%% file: finocchiaro_ckm2012_V1.tex
\pdfoutput=1

\documentclass[12pt]{article}
\usepackage{epsfig}
\usepackage{url}
\usepackage{xspace}
\def\eg{\ensuremath {e.g.}\xspace}

\def\invab   {\ensuremath{\mbox{\,ab}^{-1}}\xspace}
\def\stwob{\ensuremath{\sin\! 2 \beta   }\xspace}
\def\rhobar {\ensuremath{\overline \rho}\xspace}
\def\etabar {\ensuremath{\overline \eta}\xspace}
\def\FourS {\ensuremath{\Upsilon{(4S)}}\xspace}

\def\ASL    {\ensuremath{A_{\scriptstyle SL}}\xspace}
\def\Kz    {\ensuremath{K^0}\xspace}
\def\KS    {\ensuremath{K^0_{\scriptscriptstyle S}}\xspace}
\def\piz   {\ensuremath{\pi^0}\xspace}
\def\pip   {\ensuremath{\pi^+}\xspace}
\def\pim   {\ensuremath{\pi^-}\xspace}
\def\epem  {\ensuremath{e^+e^-}\xspace}
\newcommand{\etapr}{\ensuremath{\eta^{\prime}}\xspace}
\def\jpsi     {\ensuremath{{J\mskip -3mu/\mskip -2mu\psi\mskip 2mu}}\xspace}

\def\Bbar    {\kern 0.18em\overline{\kern -0.18em B}{}\xspace}

\def\Bz      {\ensuremath{B^0}\xspace}
\def\Bzb     {\ensuremath{\Bbar^0}\xspace}
\def\BzBzb   {\ensuremath{\Bz {\kern -0.16em \Bzb}}\xspace}
\def\Bp      {\ensuremath{B^+}\xspace}
\def\Kp      {\ensuremath{K^+}\xspace}
\def\bpsiks     {\ensuremath{\Bz \to \jpsi \KS}\xspace}

\input econfmacros.tex
\def\babar{\mbox{\sl B\hspace{-0.4em} {\small\sl A}\hspace{-0.37em} \sl B\hspace{-0.4em} {\small\sl A\hspace{-0.02em}R}}\xspace}

\def\Title#1{\begin{center} {\Large {\boldmath{\bf #1}} } \end{center}}

\begin{document}

\Title{Prospects at Future $B$ Factories}

\bigskip\bigskip

\begin{raggedright}  

{\it Giuseppe Finocchiaro\index{Finocchiaro, G.}\\
Laboratori Nazionali di Frascati\\
dell'INFN\\
Via Enrico Fermi 40, 00044 Frascati, ITALY}
\bigskip\bigskip
\end{raggedright}

Proceedings of CKM 2012, the 7th International Workshop on the
CKM Unitarity Triangle, University of Cincinnati,
USA, 28 September - 2 October 2012.

\section{Introduction}
The present flavour physics landscape has been reshaped in the last
decade by the \babar and Belle $B$ Factory experiments and by D0
and CDF at the Tevatron, and is now evolving at LHC. The CKM paradigm
has been challenged with increasing precision, but is standing all
tests so far.

The projects Super$B$ and SuperKEKB for two next-generation asymmetric
\epem flavor factories aim at collecting data samples of
$50-75$\invab, with therefore a $\times100$ increase in the statistics
collected by \babar and Belle\footnote{
At the time of writing these proceedings it was announced that the
Super$B$ project\cite{SuperB_CDR,SuperB_site} to be built at the
Nicola Cabibbo Laboratory between Frascati and Rome, Italy, was shut
down by the Italian Government due to funding problems. While the
physics of mixing and mixing-related $CP$ violation in the $B$ system
is clearly in common between the Belle II\,\cite{BelleTwo} and the
Super$B$ experiments, I will not further discuss the latter in the
remainder of the present report.}.
The Belle II experiment built at the SuperKEKB facility in Japan is
expected to start taking data in 2016, and to collect a sample of
50\,ab$^{-1}$ by 2021.  At that time the LHC experiments will have
fully explored the energy scale ${\cal O}(1\rm{TeV})$ they are
designed to have access to, and the High Energy Physics community will
face one of two possible situations.
In case signals of New Physics (NP) are found at the LHC, Belle II
could study its flavor structure, measure the flavor couplings and
search for still heavier mass states.
Alternatively, the NP energy scale could lie above the direct reach of
the LHC -- indeed, model-dependent direct searches performed with the
limited data samples collected to date have found no evidence of
physics beyond the Standard Model, SM.
In such a scenario Belle II can look for indirect NP
signals, understand where they may come from, and exclude regions in
the multi-dimensional parameter space of NP models up to
$\Lambda\simeq10\,\rm{TeV}$ or more. In addition to probing energy
scales higher than those accessible at the LHC via virtual processes,
a Super Flavor Factory can, thanks to its clean \epem environment,
experimentally access several physics channels precluded to hadronic
machines as LHC. Those channels include decays with neutrinos or many
neutral particles in the final state, or whenever an inclusive
analysis is required.

\section{\boldmath{Hunting for New Physics at a Super $B$ Factory}}
Although there is general agreement that the SM can only be the
low-energy manifestation of a more complete theory at a larger mass
scale, a variety of NP models exist, with many non predicted
parameters, including the New Physics mass scale $\Lambda_{NP}$. In
this respect, the search for NP is a complex task which can only be
accomplished with a multilateral approach, in which a high-luminosity
$B$ Factory will play a crucial role.  A recent example of the
potentiality of precision ``low energy'' measurements in constraining
NP models was discussed at this Workshop and previously presented at
the FPCP Conference\,\cite{babar_BDtaunu} with the $B\to
D^{(\ast)}\tau\nu$ decays.  The \babar\ data on the branching
fractions for these decays (normalized to the corresponding $B\to
D^{(\ast)}\ell\nu$ ones, with $\ell=e,\mu$ to reduce experimental
uncertainties) disagree with the SM at the 3.4$\sigma$ level. On the
other hand, since there is no value of the ratio $\tan\beta/m_{H^+}$
that can simultaneously accommodate the $B\to D\tau\nu$ and the $B\to
D^{\ast}\tau\nu$ measurements, these data also exclude the widely
discussed type II two-Higgs-doublet model.

As demonstrated by the several hundred papers published by the
$B$ Factory experiments, a wide range of important measurements can be
performed at the \FourS resonance. Most of these
are statistics-limited, and would therefore improve substantially with
a data sample of 50\invab. In many cases, large control samples can be
used to further reduce systematic and theoretical errors. Control of
the theoretical errors (e.g. those related to lattice calculations) is
particularly relevant to fully exploit the statistical power of the
experimental measurements.
In the last few years the global fits of flavour
data\,\cite{CKMFitter,UTFit} have at times highlighted imperfect
agreement among some of the angles or sides of the Unitarity Triangle
(UT), which is tempting to interpret as hints of physics beyond the
Standard Model. These ``tensions'' were at the 3$\sigma$ level or
less, and have always been resolved as statistical fluctuations so
far. These tests of the internal consistency of the SM will become
significantly more stringent when the experimental errors are reduced.
With 50\invab a Super $B$ Factory will also be able to substantially
improve on the precision of the CKM Unitarity Triangle
parameters. For example, the present error on the \rhobar\ and
\etabar\ parameters of the UT in a global fit where New Physics
contributions are allowed ($\pm0.056$ and $\pm0.036$ respectively),
could be reduced to $\delta\rhobar=0.005$, $\delta\rhobar=0.005$. Such
an improvement will be crucial for many NP searches with flavor, both
in the $B$ sector and elsewhere, \eg\ in the Kaon sector.

The possibility to study a very large numbers of physical observables, and
the correlations among them, is a particularly important tool to
elucidate the nature of new physics, should deviations with respect to
SM predictions be observed. 

\section{\boldmath{New Physics Contributions in $\Delta F=1$ Processes}}
The measurement of \stwob through mixing-induced $CP$ violation in the
decay \bpsiks, one of the theoretically cleanest measurements that can
be made in flavour physics, was among the main motivation for building
the current generations of $B$ Factories, and its precise measurement
(the present HFAG\,\cite{HFAG} average from \babar\ and Belle is
$\stwob=0.665\pm0.023\pm0.010$) is indeed a great success of the
factories. The precision on \stwob will be further improved at a Super
$B$ Factory. After some 10\invab the systematic error will dominate;
the theoretical error from possible penguin contributions can instead
be controlled to the desired precision using the $\Bz\to \jpsi \piz$
and $\Bp\to \jpsi \Kp$ decay channels\,\cite{CPS, FleischerCKM2012}.
The sine coefficient $S_f$ in mixing-induced $CP$ violation has the
same CKM matrix coefficients in the Standard Model for the $b\to c\bar
c s$, $b\to c\bar c d$, $b\to s\bar s s$ and $b\to d\bar d s$
modes. The latter three modes are loop-dominated processes and can
receive sizable contributions from New Physics; it is therefore
important to compare them with the reference value of $\sin2\beta$ in
$b\to c\bar c s$. This is especially relevant for some of the $b\to
s\bar s s$ decay channels such as $\Bz\to\etapr\KS$, whose
experimental signature and theoretical calculations are clean enough
to allow a precise test. A summary of the most significant of these
modes, comparing the precision at present and the one attainable at a
Super $B$ Factory is shown in Table\,\ref{Tab:DF_one}.
\begin{table}[!h]
\begin{center}
\begin{tabular}{|l|ccc|cc|}
\hline
Mode   & \multicolumn{3}{|c|}{Current Precision} 
& \multicolumn{2}{|c|}{Predicted Precision (75\invab)} \\
                                  & Stat.   & Syst.  & $\Delta S_f$ & Stat. & Syst. \\ \hline
\bpsiks    & 0.022 & 0.010 & $0\pm0.01$ & 0.002 & 0.005 \\ \hline
$\etapr\KS$ & 0.08 & 0.02 & $0.015\pm0.015$ & 0.006 & 0.005 \\
$f_0\KS$      & 0.18 & 0.04 & $0\pm0.02$ & 0.012 & 0.003 \\
$\KS\KS\KS$  & 0.19 & 0.03 & $0.02\pm0.01$ & 0.015 & 0.020 \\
$\phi\KS$      & 0.26 & 0.03 & $0.03\pm0.02$ & 0.020 & 0.005 \\ \hline
$\jpsi\piz$    & 0.21 & 0.04 &  & 0.016 & 0.005 \\ \hline
\end{tabular}
\caption{Summary of current and predicted precision for the cleanest
  $b\to s\bar s s$ modes, compared to \bpsiks, from \cite{SuperB_WP_Physics}.}
\label{Tab:DF_one}
\end{center}
\end{table}

At this workshop we have seen very beautiful analyses by the LHCb
experiment, which is clearly a competitor a high-luminosity $B$
Factory will have to deal with. In general, the well-defined initial
state and a more hermetic detector at an \epem machine allow better
tagging efficiency, better reconstruction of modes with many neutral
particles and/or neutrinos in the final state, and in some cases a
better control of systematic effects, thus counterbalancing the larger
statistical power LHCb gets from the higher hadronic cross section.
The high luminosity $B$ Factories and LHCb can therefore produce
complementary measurements to test the SM.

\section{More Physics Opportunities}
$CP$ violation in mixing, predicted in the SM\,\cite{BBLU,CFLMT} to be
$\O(10^{-3})$, has been searched for at the $B$ Factories using
same-sign dilepton events in which both $B$ mesons decay
semileptonically. The search was conducted both inclusively and in
events in which one of the $B$ decays is at least partially
reconstructed; the latter method trades some loss in statistical power
with a better control of systematics.  Still another variation of the
method can be adopted, in which the semileptonic asymmetry \ASL is
measured comparing mixed ($\ell^\pm K^\pm$) vs. unmixed ($\ell^\pm
K^\mp$) events in which the flavour of one of the $B$ mesons is tagged
by a kaon. A preliminary results using the technique above was
presented at this conference\,\cite{MargoniCKM2012}, yielding a value
of \ASL consistent with the world average, but more precise. We are
longing to see this analysis published.

Since the current \ASL measurements are already reaching the level of
being limited by systematic uncertainties, even with the Super $B$ Factory
data sample it will require a lot of effort to push down the
precision -- this will presumably require a smart data-driven
approach. But given that the SM prediction is ``around the corner'', the
potential reward of observing $CP$ violation in the mixing in the $B$
system is certainly worth the effort. In addition there is always the
possibility that physics beyond the SM is at work, and $CP$ violation
in the mixing is sensitive to such effects\,\cite{LN,LNCDJKLM,LNCDLMNT}.

A tantalizing $2.8\sigma$ effect of sidereal-time dependent $CPT$
violation using dilepton events has been around for
years\,\cite{babar_2006}, and it will be interesting to repeat such
measurements with the Super $B$ Factory data samples. Analogously, it
would be very important to repeat with 50\invab the new \babar\
measurement\,\cite{babar_2012_T_violation} which used 
$\BzBzb\to\jpsi\Kz,~\ell^\pm X^\mp \nu$ decays to perform a
simultaneous test of $T$, $CP$ and $CPT$ violation without any
a-priori assumptions.

Mixing-induced $CP$ violation in $b\to s \gamma$ transitions is
suppressed in the SM (\eg\ $|S_{\Bz\to\KS\piz\gamma}|<0.1$ even considering
strong interaction uncertainties\,\cite{GrinPir}) because the photons
carry opposite polarisations when produced from $b$ or $\bar b$
decays. A high precision measurement of this decay mode is very
sensitive to RH currents arising from NP. The current experimental
world average error,  $\delta S_{\Bz\to\KS\piz\gamma}=\pm0.20$,
dominated by the statistical uncertainty,
can be reduced to $\pm0.03$ with 50\invab. This mode, with three
photons in the final state and the $B$ decay vertex reconstructed by
the $\KS\to\pip\pim$ decay constrained to the measured beam-spot
position, can be efficiently measured only at an \epem collider;
similarly for other modes such as $\Bz\to\KS\eta\gamma$ and
$\Bz\to\KS\phi\gamma$. Other ways of probing the photon polarization
in $\Bz\to\KS\piz\gamma$ decays with very large data samples exist; an
interesting example proposed in \cite{belle2_physics} involves using
$\gamma\to\epem$ conversions in the detector material. The
distribution of the angle of the \epem plane with respect to the plane
in which the \KS and $\gamma$ lie depends on the photon left and right
polarization amplitudes. In the same reference it is estimated that a
4 sigma effect could be detected with 50\invab in case of maximal RH
currents.

\section{Summary}
A very concise overview of the possibilities offered by
high-luminosity $B$ Factories in mixing and mixing-related $CP$
violating processes in $B$ decays was given. An exhaustive discussion
of the (now stopped) Super$B$ project can be found
in\,\cite{SuperB_CDR, SuperB_WP_Physics, SuperB_WP_Detector,
  SuperB_WP_Accelerator}; the Super$B$ Technical Design report is
undergoing final review and will be published shortly. The Belle II
and SuperKEKB projects\,\cite{BelleTwo} are progressing well, and are
on track for starting the data taking in 2016.

\end{document}

%% file: econfmacros.tex
\def\beq{\begin{equation}}
\def\eeq#1{\label{#1}\end{equation}}
\def\eeqn{\end{equation}}

\def\beqa{\begin{eqnarray}}
\def\eeqa#1{\label{#1}\end{eqnarray}}
\def\eeqan{\end{eqnarray}}

\let\bar=\overbar

\def\etal{{\it et al.}}

\def\eg{{\it e.g.}}

\def\O{{\cal O}}

\def\Dslash{\not{\hbox{\kern-4pt $D$}}}
\def\dslash{\not{\hbox{\kern-2pt $\del$}}}

\def\msb{{\bar{\ssstyle M \kern -1pt S}}}

